\documentclass[12pt]{article}
\usepackage{epsfig,graphics,color}
\author{\large \bf  Z.~Usubov\footnote
         {On leave of absence from Institute of Physics, Baku, Azerbaijan}
\\
\\Joint Institute for Nuclear Research,
\\ Dubna, Russia}          

\title { Looking for Large Extra Dimensions in the \\      
                           Early LHC Data }              

\begin{document}
\maketitle
{
\vskip 0.5cm
\hskip 6.5cm {\bf \large { Abstract}}
\vskip 1.0cm

We explore the opportunity to look for large extra dimensions 
in the early stages of the LHC running. The high-E$_T$
dijet production is analyzed via a novel kinematic variable 
when the machine center-of-mass energy varies from 7 TeV through
10 TeV to 14 TeV and the accumulated data range from 0.5 to 8 fb$^{-1}$.
The estimations of the reach in the effective scale $M_S$ 
for different numbers of large extra dimensions 
are presented.
\vskip 0.4cm


\large {
\section{Introduction}
$ $

There is  a common belief  that the Large Hadron Collider (LHC)
will provide a key insight into the electroweak symmetry breaking
in the Standard Model (SM).
Even if the Higgs sector will be established by the Tevatron 
 or forthcoming LHC     
experiments we will still have few fundamental motivations for physics            
beyond the SM: the existence of dark matter and dark energy, flavor 
mixing and CP violation, quadratic mass divergences in scalar sector,
unified descriptions of the SM and quantum theory of gravity, baryon     
asymmetry, hierarchy problem, neutrino masses, etc. 

The main efforts for completely formulated models beyond SM
are dedicated to the supersymmetry - the highway
of new physics expectations.
One of the possible avenues  beyond the SM  is  the models
with large extra dimensions\footnote {
The existence of extra spatial dimensions is the main feature of string
theories in which  the  characteristic size is of order of the Planck
length, $1.6 \times 10^{-33}$ cm},
namely the models in which
our space-time is not 3+1-dimensional at all. These  models
offer the possibility of reducing  the real gravity scale to a value as        
small as $\cal O$(1 TeV). 

In the model\cite{add} the authors add $N_{ED}$ spatial
extra flat dimensions to the space-time structure, i.e. our world, 
3+1-dimensional  brane,  is embedded in a higher-dimensional structure, 
bulk. All SM particles are confined in the brane, and only for gravitons
$N_{ED}$ dimensions of the bulk are transparent. Thus,      the       
extra dimensions would be probed only via graviton interactions.

Another possible extension of space-time\cite{rans} is a                     
5-dimensional bulk with a 
nonfactorizable geometry. The only single extra dimension, finite
or infinite, is warped by an exponential factor. In this case the
effective Planck scale varies from $10^{19}$ GeV to few TeV across
the 5th dimension.

In the model with the universal extra dimension\cite{acd} the extra space                 
coordinate is accessed by all SM particles. One of the features of 
this model is the possibility of  providing  interesting dark matter
candidates: the lightest Kaluza-Klein photons and neutrinos\cite{serv}.           
 
The size of extra dimensions  in various theories 
varies in a wide range  from $\cal O$(1 fm) to infinity. In the latter
case they are hidden.

After August 2009 CERN announcement the particle physicists  pin                 
their hopes to the 3.5 TeV per beam run in  2009-2010.
Of course, the early objects of the study will be commissioning
and operation stability of the accelerator and detectors.
After  the "rediscovery" of the SM -- W, Z, J/$\psi,\,\Upsilon,\,
\tau,\,t\bar t,\,2-jet$, etc. will be produced copiously at the LHC --
the search for physics beyond it   will begin.
Signatures of new physics at the TeV scale will obviously be probed          
in parallel with these efforts. 

The rest of this note is organized as follows. The  next section
gives a brief description of flat extra dimensions as 
introduced in\cite{add}. Section~3 describes our strategy for the 
search for
large extra dimensions  at the LHC.
In Section~4 we present  the details of our simulation
and analysis technique. Section~5 is  the  summary        
of our analysis itself. We examine large extra dimensions       
with the novel kinematic variable. The sensitivity of the results
to the choice of the parton distribution functions and energy
smearing in the hadron calorimeter is  also demonstrated.
In this Section we estimate the LHC reach for effective energy scale of      
the large extra dimensions.
We end with the conclusions in Section~6.

\section { One Possible Extension of Space-Time Dimensions}
$  $

We will follow the theoretical framework proposed  
in\cite{add} (hereafter ADD). In this scenario
the relation between the effective Planck scale for the bulk $(M_{eff})$
and for the 3+1-dimensional brane $(M_{Pl})$ is governed by the equation 
\begin{equation}
 M_{Pl}^2 = 8 \pi M_{eff}^{N_{ED}+2} R^{N_{ED}},
\end{equation}
where all $N_{ED}$ extra dimensions are compactified to a radius $R$.
Note  that if one puts $M_{eff}\sim \cal O$(1 TeV) to avoid the hierarchy 
problem, $R$ becomes very large for $N_{ED}=1\,(R \sim 10^8$km) and
varies from $\sim 0.1$ mm to a few fm when $N_{ED}$ ranges from 2 to 7.
This speculation leads to the distortion of the usual inverse square
law of gravity at $r <  R$.                                              
Experimentally allowed values for the compactified radius of extra 
dimensions must be smaller  than  44 $\mu$m\cite{kea}.

The addition  of extra dimensions leads  to numerous excited  
states for the particles living in the bulk. The tower of
graviton states with nonzero momentum  which couple to
SM particles is called Kaluza-Klein (KK) excitation. The mass
spectrum of graviton KK states is given by $m_l^2=l^2/R^2$, where       
$l=1,2,3...$. The universal coupling is obtained by summing over
all the KK states, which leads to the strength of interactions with
SM particles of the order of $1/M_{eff}$. Thus, direct emission of gravitons    
or effects caused by virtual exchanges of KK states will be detectable
in SM particle interactions at accessible energies.

Comprehensive investigation of 
graviton-involving subprocesses was done                 
in \cite{grw1,hlz,hew}.                                  
It  was  shown that  virtual graviton 
exchange effects are
sensitive to the ultraviolet cutoff $M_S \sim {\cal O}(M_{eff})$, which is 
necessary to keep the sum  over  KK states nondivergent.
If energy scale is above $M_S$, the string dynamics should be taken into account.
The cross section of $2 \to 2$ subprocesses in the presence of large extra
dimensions was parametrized by the variable $\eta  ={\cal F}/M^4_S$: the
pure graviton exchange part is quadratic in $\eta  $, and the interference         
one is linear in $\eta  $. Different formalisms lead to the 
following definitions of {$\cal F$}:
\begin{equation}
{\cal F}=1,\hspace{6.25cm} \cite{grw1}                    
\end{equation}
\vspace{0.0cm}
\begin{equation}
{\cal F}= { \Biggl\{ { log{({M_S^2\over \hat s})}\atop{2\over{N_{ED}-2}}} } 
\qquad {N_{ED}=2\atop N_{ED}>2,}\qquad\quad\,\cite{hlz}
\end{equation}
\vspace{0.20cm}
\begin{equation}
{\cal F}={2\lambda \over \pi}={\pm {2\over{\pi}},}{\hspace{4.8cm}}\cite{hew}
\vspace{0.3cm}
\end{equation}
Obviously, the SM prediction is recovered in the limit $M_S \to \infty$.

The LEP and Tevatron Collaborations have intensively searched for direct 
graviton productions in $e^+e^-$ and $p \bar p$ interactions. The combined
LEP limits for $M_S$ are $M_S>1.6$ TeV  for $N_{ED}=2$ and $M_S>0.66$ TeV          
for $N_{ED}=6$ at the $95\,\%$ CL \cite{leplim}. 
The CDF and D0 Collaborations have looked for large extra dimensions using  
different channels\cite{d0cdf,d01}. The best limits on $M_S$ are 2.09--1.29 TeV
at the 95$\%$ CL for $N_{ED}=2-7$,    obtained by D0\cite{d01} using dielectron
and diphoton channels within the          formalisms\cite{hlz}.
 
\section{The  Strategy for Extra Dimension Study at the LHC}
$ $

The influence  of the virtual graviton exchange on the dijet 
production in  proton-proton 
collisions at the  ATLAS  may be a promising hint to the study of 
extra dimensional gravity scenarios.   
A typical signature would be a more isotropic dijet angular
distribution than expected from the SM 
predictions and/or excess of the high-$E_T$ jets over the level 
predicted by QCD.
The angular distribution of the jets is  
sensitive to the new physics and less susceptible to the
systematic uncertainties\cite{ptdr}.
The dijet angular distribution becomes especially interesting  because 
it could  reflect the  spin-2 nature of the gravitons.

The analysis of dijet angular distribution is quite often based on the 
variable  $\chi$\cite{barfil}                

\begin{equation}
   \chi = { \hat u \over \hat t } = {exp {\left| ({\eta}_1 - {\eta}_2) \right|}}
    = { {1+ \left| cos {\theta}^* \right|} \over {1 - \left| cos {\theta}^* \right|}}, 
\end{equation}
where $\hat u$, $\hat t$ are the usual Mandelstam variables for $2\to 2$ 
subprocesses, ${\eta}_{1,2}$ are the pseudorapidities of the 
leading jets, ${\theta}^*$ is the center-of-mass scattering angle.

Earlier  we defined the new kinematic variable\cite{usu1} for the high-E$_{T}$      
dijet   final state and showed that this variable was very useful for
analysis of quark compositeness at the LHC. We define the           
dimensionless variable  $\alpha_{Z}$ in $pp$ interactions at center-of-mass 
energy E$_{CM}$ as $E_{CM} \over 2$ times the 
ratio of the sum  and the product of the                                           
transverse momenta of two hardest jets 
\begin{equation}
   \alpha_{Z} \equiv { E_{CM}  \over 2} \,\,{  { P_{T_1}+P_{T_2} } 
            \over { P_{T_1}\times P_{T_2}}}.
\end{equation}  

 The estimation of the effective energy scale $M_S$ in this study
is based on the analysis of
${\alpha}_Z$. To describe the whole distribution with a                     
single parameter we consider the variable
\begin{equation}
   R_{{\alpha}_{Z}} ={N({\alpha}_{Z}<{\alpha}_{Z}^{0}) \over
                N({\alpha}_{Z}>{\alpha}_{Z}^{0})},       
\end{equation}
where $N({\alpha}_{Z} > {\alpha}_{Z}^{0})$ 
($N({\alpha}_{Z} < {\alpha}_{Z}^{0})$) 
is the number of 
dijet events with ${\alpha}_{Z} > {\alpha}_{Z}^0$     
(${\alpha}_{Z} < {\alpha}_{Z}^0$).        

In order to know to what extent the observations   
either  conform or disprove the  spatial large extra dimensions 
scenario we consider the significance
\begin{equation}
   S = {{ \left|R_{{\alpha}_Z}(ED) - R_{{\alpha}_Z}(SM) \right|} 
              \over {\sigma}},    
\end{equation}
where $\sigma$ is calculated as the sum in quadrature of ${\sigma}_{SM}$
and ${\sigma}_{ED}$.
               
\vskip  0.6cm
\begin{figure}[ht]
\vskip -0.3cm
\centerline{\epsfxsize 5.0 truein \epsfbox{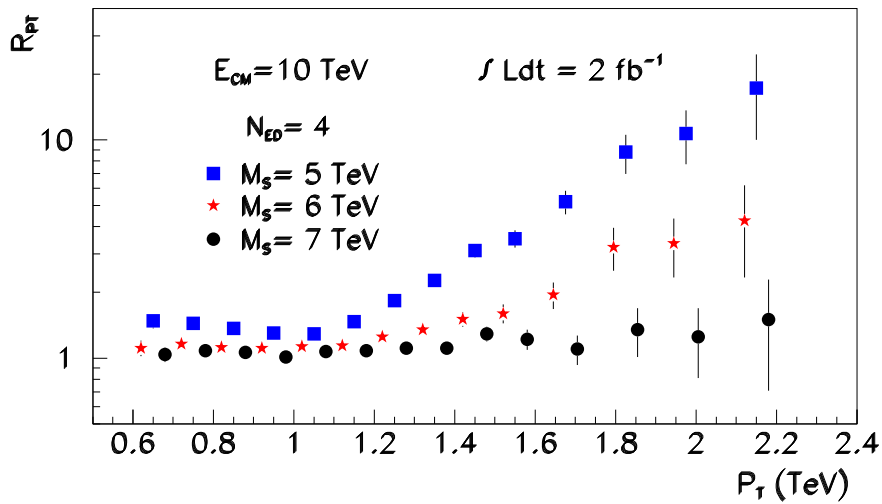}}
\vskip -6.0cm
\caption{ The ratio of the $P_T$ distribution of two leading jets
in the model with large extra dimensions 
to the one predicted by the  Standard Model.}                                  
\end{figure}
\newpage
\section{Data simulation for the generic LHC detector}
$ $

The simulation of the $pp$ collision was performed
with the event generator PYTHIA6.4\cite{pyth}.              
The parton-parton cross sections at the tree level 
including the effects from 
off-shell gravitons and their interference with the SM 
were  taken from\cite{atwo} and  properly
incorporated in PYTHIA.
We employ the leading-order CTEQ6L1\cite{pump}  pdfs everywhere 
unless  otherwise stated and use PYTHIA6.4
default choices for $Q^2$ definition as well as
factorization/renormalization scales.
The initial-state and final-state 
QCD and QED radiation and multiple interactions were enabled.
The events were generated with the hard subprocess transverse      
momentum  $p_{T} > 1\,$TeV. 

\begin{figure}[ht]
\vskip -0.3cm
\centerline{\epsfxsize 5.0 truein \epsfbox{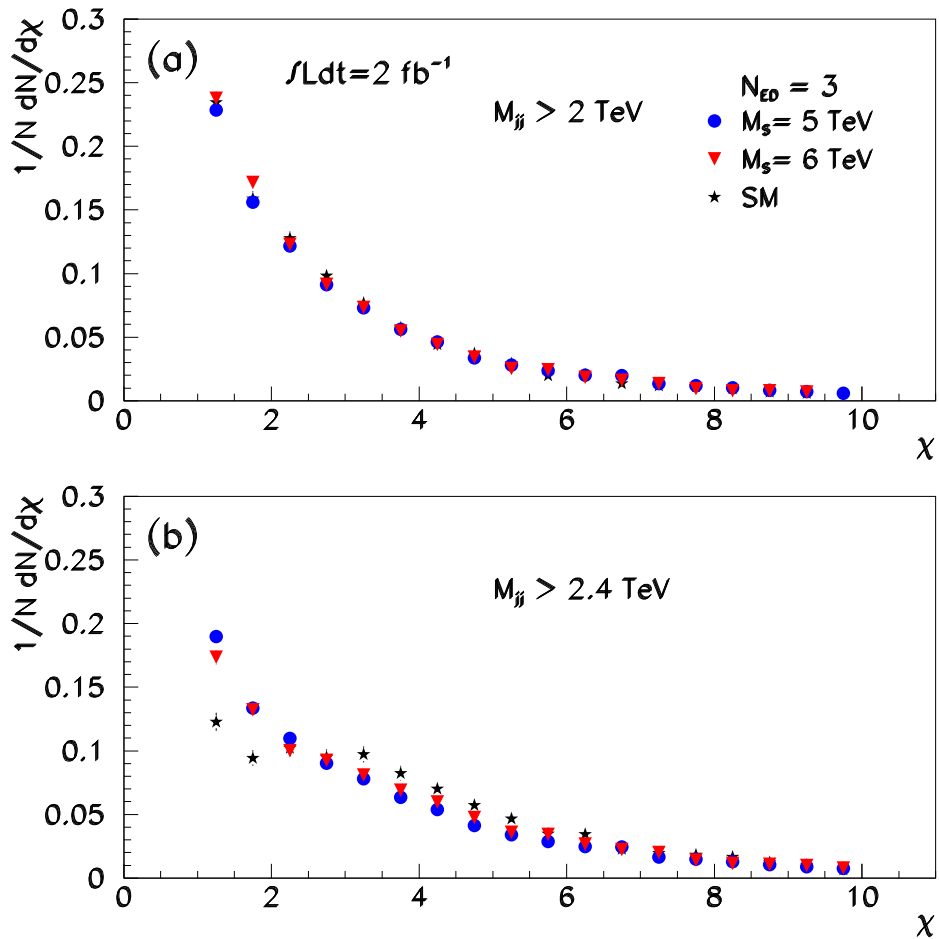}}
\caption{The Standard Model prediction for the dijet
angular distribution 
compared to the model with large extra dimensions expectations 
at different energy scales $M_S$:
$(a)$ the dijet invariant mass $M_{jj}>2$ TeV;
$(b)$ $M_{jj}>2.4$ TeV. 
The integrated luminosity is assumed to be 2~fb$^{-1}$ and $E_{CM}=10$~TeV.} 
\end{figure}

The detector performance was simulated by using the 
publicly available PGS-4\cite{conw} 
package written by J.~Conway and modified
by S.~ Mrenna for  the generic  LHC detector.                         
The calorimeter granularity is set to 
$(\Delta \phi \times \Delta \eta)=(0.10 \times 0.10)$. Energy smearing
in the hadronic calorimeter of the generic LHC detector is governed by\footnote
{We add                            
in the PGS-4 simulation of energy smearing
in the hadronic calorimeter the constant term }
\begin{equation}
  {\Delta E \over {E}} = {{ a  \over \sqrt{E}} \oplus b   } \qquad (E\,\,in\,\,GeV),
\end{equation}
where the stochastic term factor is $a=0.8$ and the 
constant factor is  $b=0.03$.
Jets were reconstructed down to $|\eta|\le 3$ using 
the  $k_T$ algorithm implemented in PGS-4. We chose $ D=0.7$
for the jet resolution parameter and required that both leading jets 
carried a transverse momentum $P_{T}^{j1,j2}>100\,$GeV.
We use the simplified     
output from PGS-4, namely, a list of two most energetic
jets.
The average $P_T$
and invariant mass of the leading jets are
$<P_T^{j1}>=1.12\,$TeV, $<P_T^{j2}>=1.01\,$TeV, 
$<m_{j1j2}>=2.46\,$TeV, respectively.

\section{Effect of Large Extra Dimensions on Early LHC Data}
$ $
 
Even in the early stages of its operation, the LHC allows one to reach
very large values of jet transverse energy and dijet invariant
mass. This   kinematic region of dijet production has never been
studied before.
\begin{figure}[ht]
\vskip -0.3cm
\centerline{\epsfxsize 5.0 truein \epsfbox{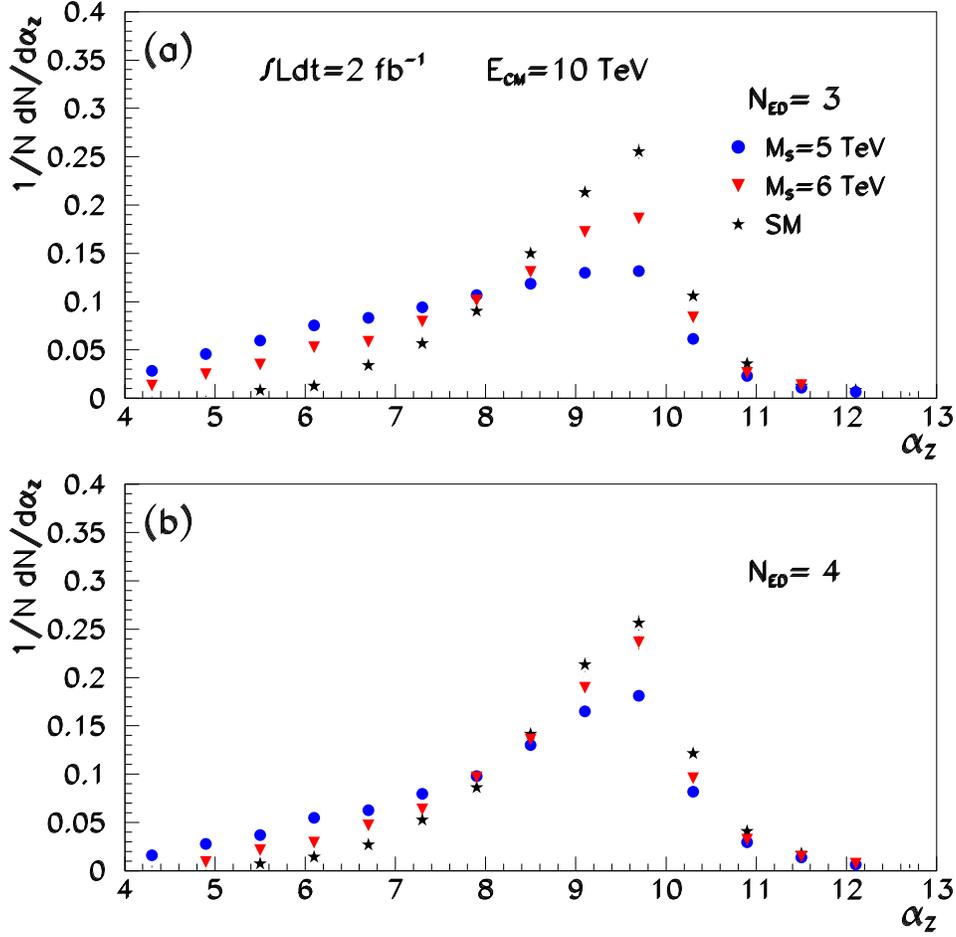}}
\caption{                                                               
The Standard Model prediction for the normalized ${\alpha}_{Z}$ distribution
(see the text)                  
compared to the predictions of the model with large extra                  
dimensions at different energy scales $M_S$:                              
$(a)$  the number of compactified extra dimensions $N_{ED}=3$;
$(b)$  $N_{ED}=4$.                          
The integrated luminosity is assumed to be 2 fb$^{-1}$ and $E_{CM}=10$ TeV.} 
\end{figure}

The model with large extra dimensions, as  has 
already been noted, can manifest itself    through the
deviation of the jet transverse momentum and/or dijet angular
distributions from the QCD prediction.
In Fig.~1 we plot the ratio of the $P_T$
distribution of the two hardest jets, $R_{PT}$,
derived in the ADD model with different $M_S$ to the
$P_T$ distribution predicted by the SM.
The number of extra dimensions $N_{ED}$
was chosen to be four. 
The enhancement of the ADD model
cross section over the SM prediction is obvious  at high values 
of jet $P_T$. Figure~1 as well as Figs.~2,3   hereafter examine
the effects of large extra dimensions at $E_{CM}=10$ TeV and the
shown sensitivity corresponds to the integrated luminosity of
2 fb$^{-1}$.                                          

A comparison  between the dijet angular distribution predicted by the SM
and induced by the ADD model                                                          
is shown in Fig.~2. Plots $(a)$ and $(b)$ correspond to the
normalized $\chi$ distributions, $(1/N)(d N/d \chi)$, for two values 
of $M_S$ and two lower limits for $M_{jj}$, 2.0 and 2.4 TeV respectively.
The data were obtained  at   $N_{ED}=3$.                                                       
Obviously,
the effect of extra dimensions is most pronounced in the high
dijet mass region.

\begin{figure}[ht]
\centerline{\epsfxsize 5.0 truein \epsfbox{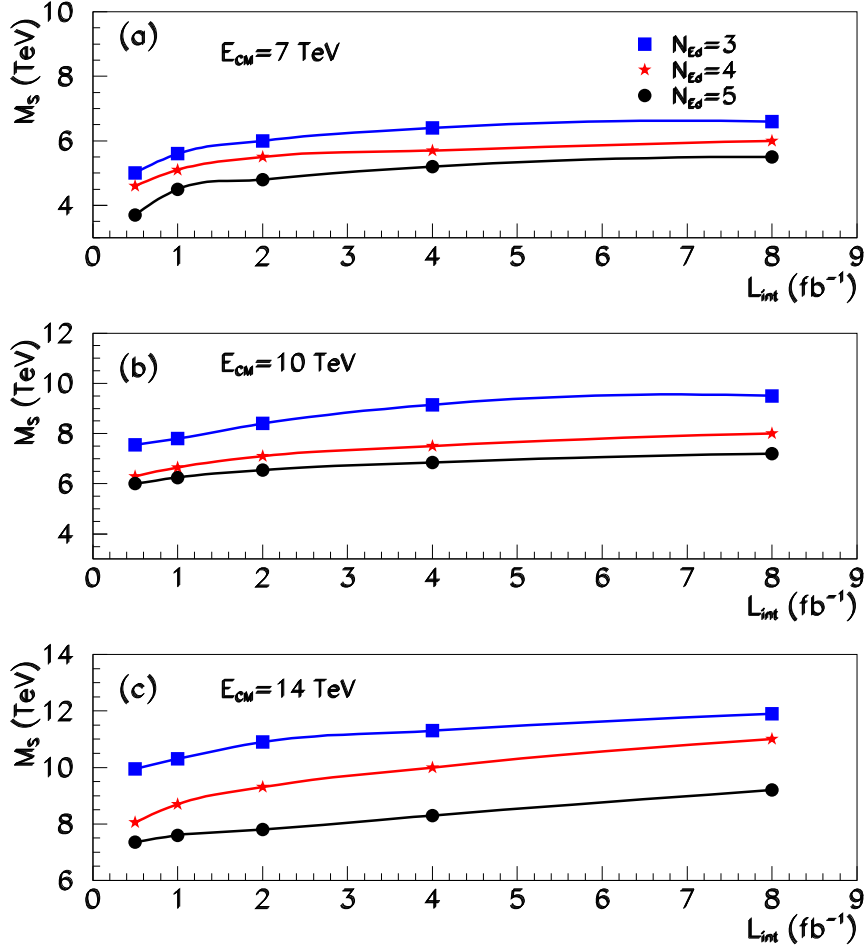}}
\caption{Expected LHC reach on the effective energy scale $M_S$ as a       
function of the collider  integrated luminosity for different               
numbers of extra dimensions $N_{ED}$: 
$(a)$ $E_{CM}=7$ TeV;              
$(b)$~$E_{CM}=10$ TeV and $(c)$ $E_{CM}=14$ TeV.} 
\end{figure}

The normalized ${\alpha}_Z$ distributions predicted by the SM
and the ADD model for $N_{ED}=3$ and 4
are demonstrated in Fig.~3$(a)$ and Fig.~3$(b)$, respectively. 
Various points  correspond to different values of $M_S$.
Compared to the $\chi$ distributions of Fig.~2$(a)$, 
we have a robust signal for ${\alpha}_Z$ from large extra dimensions              
at $M_{jj}>2$ TeV already. The net result is that the significance            
as defined in Eq.(8) 
for ${\alpha}_Z$ at $M_{jj}>2$ TeV is more than 15(13) times larger          
than that for the $\chi$ distribution for $M_S$=5(6) TeV.                    
As becomes apparent from the figure, extra dimensions lead to    
enhancement (abatement) of the distributions at
${\alpha}_Z < 8.0\,({\alpha}_Z > 8.0)$ in comparison to the 
SM prediction. 

\begin{figure}[ht]
\centerline{\epsfxsize 5.5 truein \epsfbox{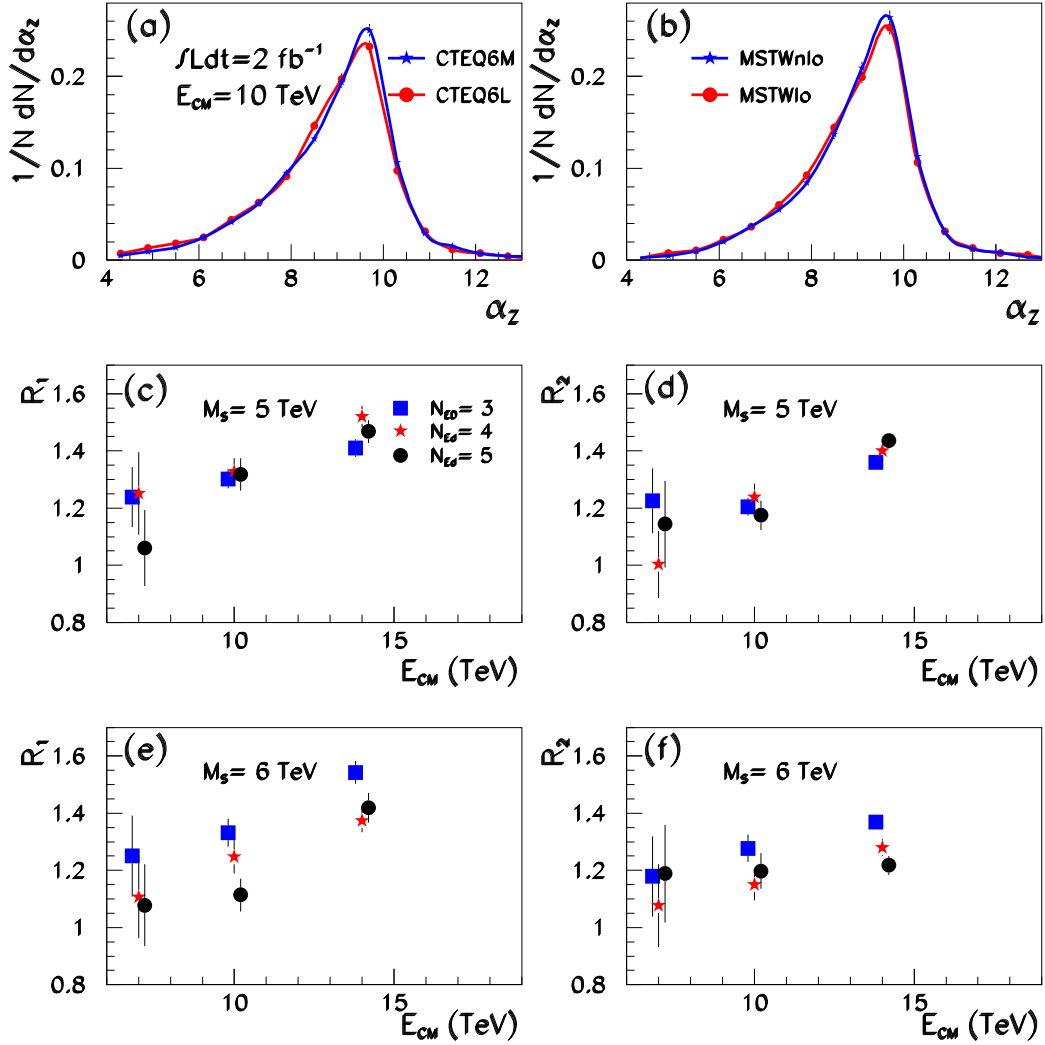}}
\caption{The influence of the choice of the parton distribution functions 
on  ${\alpha}_{Z}$ and $R_{{\alpha}_Z}$ (see the text for details).}           
\end{figure}

The sensitivity of the future LHC experiments        to the 
parameter $M_S$ up to which the effects of large extra dimensions
can be observed for $E_{CM}=7$, 10 and 14 TeV are  summarized in
Figs.~4$(a)$, 4$(b)$ and 4$(c)$, respectively.
The corresponding $M_S$ reach is shown for $N_{ED}$=3, 4 and 5 as a
function of the accumulated luminosity.
The data were  obtained with the significance as defined in Eq.(8)
close to $S=3$, for which we can claim that we have strong 
evidence for the observed signal. The parameters ${\alpha}_{Z}^{0}$
used for $E_{CM}=7,\,10$ and 14 TeV are ${\alpha}_{Z}^{0}=6.1,\,8.2$      
and 10, respectively. The calculations were done with the inclusion of
statistical uncertainties alone.

\begin{figure}[ht]
\centerline{\epsfxsize 5.0 truein \epsfbox{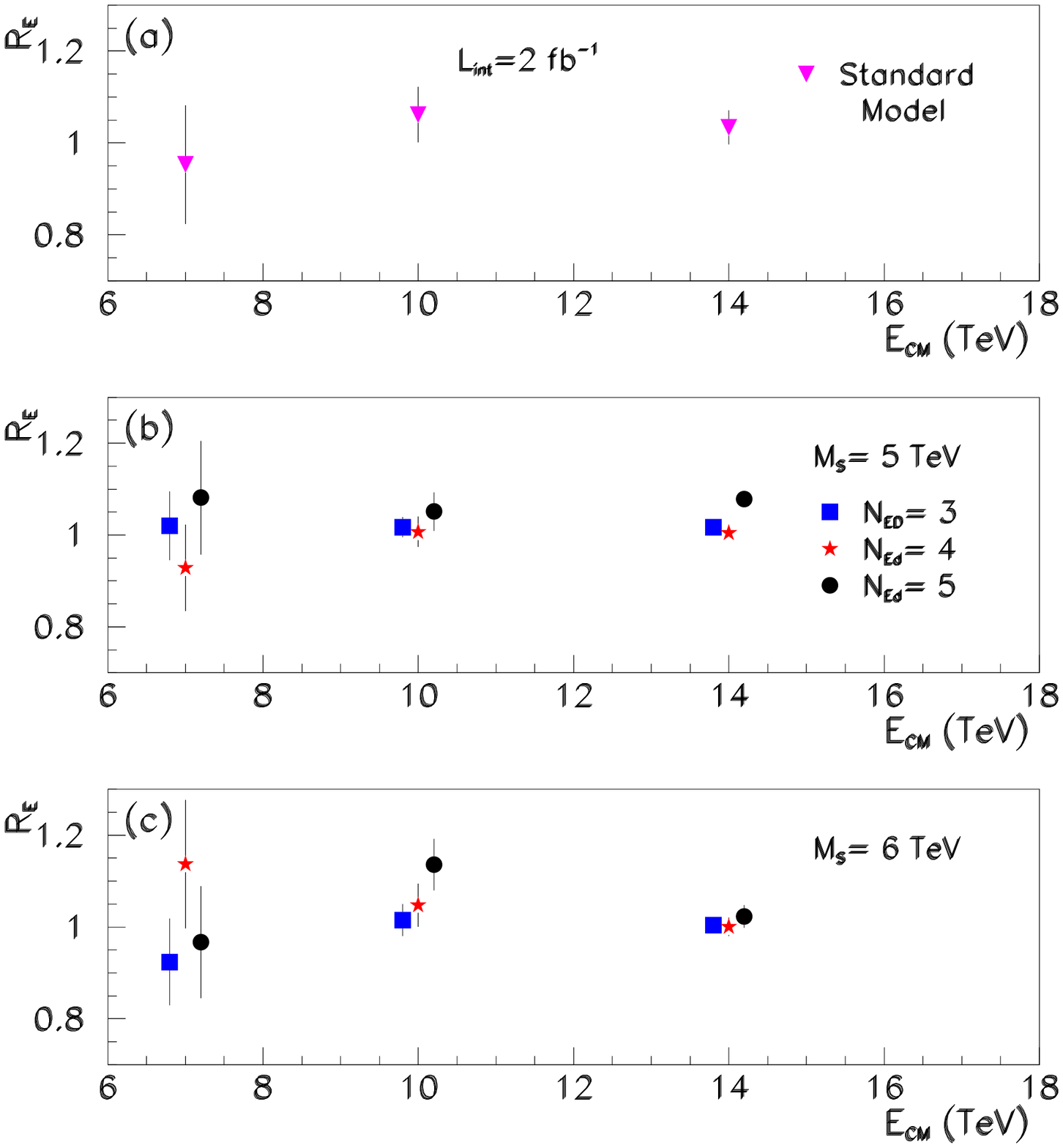}}
\caption{The influence of the hadron calorimeter energy smearing 
to the $R_{{\alpha}_Z}$ (see the text): 
$(a)$ Standard Model;
$(b)$  model with large extra dimensions, $M_S=5$ TeV;
$(c)$  $M_S=6$ TeV. 
The sensitivity correspond to the 
2 fb$^{-1}$ of data at  
$E_{CM}=10$ TeV.}                                      
\end{figure}

One of the main uncertainties in the interpretation of new          
physics at the LHC detectors will come from parton distribution                 
functions (pdfs). Unfortunately, the pdfs
are not so well determined in the kinematic region with high
transverse momentum jets. In order to show the effect of the choice
of pdfs we plot in Figs.~5$(a),\,(b)$
the normalized ${\alpha}_Z$ distribution obtained with the 
leading order (LO) and next-to-leading order (NLO) CTEQ6\cite{pump}
and MSTW2008\cite{mart} pdfs, respectively.
Shown is  the prediction  of the ADD
model with $M_S=6$ TeV and $N_{ED}=4$.
The ratios of the $R_{{\alpha}_Z}$       
obtained with the LO~(NLO)  CTEQ6  to    the one
obtained with the LO~(NLO)  MSTW2008,                            
$$ R_1 = {R_{{\alpha}_Z}(CTEQ6L1) 
\over R_{{\alpha}_Z}(MSTW2008lo)} \quad and \quad  
   R_2 = {R_{{\alpha}_Z}(CTEQ6M1) 
\over R_{{\alpha}_Z}(MSTW2008nlo)},$$
are demonstrated in Figs.~5$(c),\,(e)$ (Figs.~5$(d),\,(f)$).     
The ratios are shown  as a function of $E_{CM}$ for                
2 fb$^{-1}$ of data.                                       
Figures~5$(c),(d)$ and $(e),(f)$
correspond to the predictions of the ADD models 
with $M_S$=5 and 6 TeV, respectively.
As can be seen in these figures, $R_{{\alpha}_Z}$ 
shows  significant dependence on the choice of  pdfs. The       
ratios used also depend on $E_{CM}$.
In view of pdfs used       
we can conclude that the estimations of $M_S$ can be 
affected by the choice of  pdfs. On the other hand,                 
${\alpha}_Z$ measurements at the LHC would be used to constrain
the proton pdfs.

To examine another source of the uncertainties 
coming from hadron calorimeter energy
smearing, we show in Fig.~6   
the ratio of  $R_{{\alpha}_Z}$ derived with the
stochastic term factor in                                                
Eq.(9) $a$=1.0 to the one derived with $a$=0.5 
$$ R_E = {R_{{\alpha}_Z}(a=1.00) 
\over R_{{\alpha}_Z}(a=0.5)}.$$ 
Figure~6$(a)$  illustrates the dependence of above ratio
on $E_{CM}$ for the SM prediction, and  Figs.~6$(b),(c)$             
illustrate the one for the  ADD model.
Figure~6   exhibits  rather low        
sensitivity of $R_{{\alpha}_Z}$ to stochastic term factor $a$ 
in the range from 0.5 to 1.

\section{Conclusions }
$ $

We examined the capability of the LHC experiments to observe
large extra dimensions in the early stages of the running 
considering that the LHC would have  accumulated
luminosity in the range from  0.5 to 8.0 fb$^{-1}$ and 
$E_{CM}$ varies from 7 TeV through 10 TeV to 14 TeV.
 The distribution of the 
novel variable ${\alpha}_Z$\cite{usu1} might provide direct               
hint to observation  of 
the extra dimensions as well as dijet
transverse momentum and angular distributions.

The variable ${\alpha}_Z$ is more
sensitive to the influence of large extra dimensions and 
can be effective for estimation of the energy scale $M_S$.
This variable can show robust signatures for new physics 
even with low integrated luminosity.

Note that  collider signatures of different new physics scenarios
often mimic each others.
The use   of different kinematic variables or combinations of
them can offer an important tool for  discriminating between
new physics models.                                                      

{ \large        {
      
}}
}} 

\end{document}